\documentclass[final,3p,times,twocolumn]{elsarticle}
\usepackage{amsmath,amssymb,graphicx}
\usepackage{color}
\usepackage{multirow}
\usepackage{float}
\usepackage{subfigure}
\usepackage{rotating}
\usepackage{wasysym}

\pdfoutput=1

\newcommand{\nc}{\newcommand}
\nc{\postscript}[2]
{\setlength{\epsfxsize}{#2\hsize}\centerline{\epsfbox{#1}}}
\nc{\non}{\nonumber}
\nc{\hc}{\hbox {h.c.}} \nc{\re}{\hbox {Re}} 
\nc{\mev}{\hbox {MeV}} \nc{\gev}{\;\hbox {GeV}} \nc{\tev}{\;\hbox {TeV}}
\def\lsim{\mathrel{\raise.3ex\hbox{$<$\kern-.75em\lower1ex\hbox{$\sim$}}}}
\def\gsim{\mathrel{\raise.3ex\hbox{$>$\kern-.75em\lower1ex\hbox{$\sim$}}}}

\nc{\etal}{{\it et al.}}
\nc{\Lsp}{\;\;\;\;\;\;\;\;\;\;}  \nc{\LLLsp}{\lspace \lspace}
\nc{\lsp}{\;\;\;\;\;\;}
\nc{\spac}{\;\;\;}
\nc{\noi}{\noindent}
\nc{\beq}{\begin{equation}}   \nc{\eeq}{\end{equation}}
\nc{\bea}{\begin{eqnarray}}   \nc{\eea}{\end{eqnarray}}
\nc{\baa}{\begin{array}}      \nc{\eaa}{\end{array}}
\nc{\bit}{\begin{itemize}}    \nc{\eit}{\end{itemize}}
\nc{\ben}{\begin{enumerate}}  \nc{\een}{\end{enumerate}}
\nc{\bce}{\begin{center}}     \nc{\ece}{\end{center}}

\def\sq2{\sqrt{2}}

\def\ph{\varphi}

\def\m4{m^4(\ph)}
\def\mn2{m_n^2}

\def\v5{V^{(5)}}

\def\baa{\begin{array}}
\def\eaa{\end{array}}

\journal{Physics Letters B}

\begin{document}

\begin{frontmatter}


\title {Dark Matter from Conformal Sectors}

\vskip 20pt

\author[label1]{Durmu\c{s} Ali Demir}
\ead{demir@physics.iztech.edu.tr}
\author[label2]{Mariana Frank}
\ead{mariana.frank@concordia.ca}
\author[label1]{Beste Korutlu}
\ead{bestekorutlu@iyte.edu.tr}
\address[label1]{Department of Physics, \.{I}zmir Institute of Technology, IZTECH, TR35430, \.{I}zmir, Turkey}
\address[label2]{Department of Physics, Concordia University,
7141 Sherbrooke St. West,
Montreal, Qc., H4B 1R6, Canada}

\date{\today}

\begin{abstract}
We show that a conformal-invariant dark sector, interacting conformally with the
Standard Model (SM) fields through the Higgs portal, provides a viable framework where
cold dark matter (CDM) and invisible Higgs decays can be addressed concurrently.
Conformal symmetry naturally subsumes the ${\mathbb{Z}}_2$ symmetry needed for stability
of the CDM. It also guarantees that the weaker the couplings of the dark sector fields
to the SM Higgs field, the smaller the masses they acquire through electroweak breaking.
The model comfortably satisfies the bounds from Large Hadron Collider (LHC) and
Planck Space Telescope (PLANCK 2013).
 \end{abstract}

\end{frontmatter}

\section{Introduction}
\label{sec:intro}
As the fundamental scalar discovered at the LHC \cite{Aad:2012tfa}, highly likely to be the Higgs boson of the Standard Model (SM), 
has been the only new particle discovered so far in searches extending well above a ${\rm TeV}$, the emerging picture of 
the electroweak scale is converging to the SM, within uncertainties in determinations of Higgs boson couplings. However, 
this SM-only picture, among other vital problems like unnaturalness, suffers from having no candidate particle for cold dark matter (CDM), 
which is now widely believed to make up the bulk mass of the Universe. If CDM is to be explained by a fundamental particle, then the 
crystallizing SM-only picture must  be supplemented at least by a CDM candidate. Despite the current developments in both direct and indirect detection experiments, 
and progress in observational cosmology, understanding the particle nature of dark matter (DM), its properties and symmetries, 
and a model accommodating it, have remained elusive. To begin building a particle physics model for DM, it is important to note that:
\bit
\item The latest results on cosmological parameters, interpreted in the $\Lambda$CDM model, reveal that CDM forms $26.8\%$ of total mass in the Universe \cite{Ade:2013uln},
\item The latest LHC results on particles beyond the SM, interpreted mainly in  supersymmetry (see \cite{Jungman:1995df} for a review) and extra dimensions (see \cite{Hooper:2007qk} for a review) reveal no significant excess in processes with missing energy (plausibly taken away by the CDM particle),
\item It is thus conceivable that the CDM particle can be nestled far below the weak scale provided that its couplings to the SM spectrum are 
sufficiently suppressed.
\eit
In view of these properties, 
in the present work, we build a conservative CDM model by modifying the SM  in a minimal way, and observing that:
\bit
\item A lightweight CDM sector naturally arises if it derives from a conformal-invariant dark sector that couples conformally to the SM particles.
The reason is that all the scales in the dark sector, the CDM mass in particular, are directly generated by electroweak breaking, and, in general, the 
smaller its couplings to the Higgs field, the lighter the CDM particle. 

\item Conformal symmetry naturally accommodates the ${\mathbb{Z}}_2$ symmetry required for longevity of the CDM particle. This feature becomes
transparent especially for singlet scalars coupling to the SM Higgs field. 
\eit
In what follows, we shall construct the CDM model explicitly and analyze it against the latest results form Planck and LHC. 

Classical conformal symmetry, entering as an ideal tool into our approach to CDM, plays an important role in various other aspects of the
SM and physics beyond it. Basically, conformal symmetry forbids all fixed scales in a theory, and hence, small scales like the Higgs
mass-squared might be understood by conformal breaking. The stability of Higgs mass against quadratic divergences requires a large fine-tuning 
at  each order of perturbation theory \cite{Susskind:1978ms}, triggering a wide range of beyond the SM extensions. The desire to avoid 
such unnatural fine tuning has been the major motivation behind numerous beyond-the-SM scenarios. Among them, the conformal 
symmetry has long been considered as the symmetry principle behind naturalness \cite{Bardeen:1995kv}. It has been shown in \cite{Hur:2007uz} 
that, in a classically conformal symmetric extension of the SM, with a new hidden QCD-like strongly interacting sector, it is possible that all 
the mass scales both in the SM and in the hidden sector arise through a dynamically generated scale in the hidden sector. In this model, the connection
 of the hidden sector to the SM is provided by a messenger real singlet scalar, which then triggers spontaneous breaking of the electroweak 
symmetry of the SM. By the same token, it has been shown in \cite{Demir:2012nd} that, although quantum effects break the conformal symmetry explicitly,  
conformal duality provide a viable renormalization programme for Higgs sector. Attempts at model building in this direction had already noted that, 
in the post-Higgs era, it is preferable to consider conformal-invariant extensions of the SM. (See also the recent attempt \cite{Farzinnia:2013pga} using
conformal-invariant interactions with Coleman-Weinberg effective potential, where quadratic and quartic divergences are blinded by 
dimensional regularization scheme.)

In this Letter, we proceed based on the hypothesis that conformal symmetry automatically induces the required ${\mathbb{Z}}_2$ symmetry 
for stabilizing the CDM, and that at the classical level, it is essential for the existence of small mass scales in nature. We thus consider 
a generic, conformally-invariant DS involving scalars, gauge fields and fermions in addition to the SM particle spectrum. Each of these fields 
can be a CDM candidate depending on the symmetries of the DS. These features ensure that
a conformal-invariant DS can yield a simple and transparent model of CDM. Imposing conformal symmetry on DS provides
a naturally light, weakly interacting dark sector. The mass-squared of the SM Higgs field, the only parameter that breaks
conformal symmetry explicitly, generates all the particle masses in the SM and DS. The CDM candidate(s) acquire  mass only from its coupling through the Higgs portal, 
and the smaller the coupling of the DM to Higgs, the smaller its mass compared to electroweak scale. Conformal invariance enhances the predictive
power of the model, and numerical analysis shows that conformal coupling of DS to Higgs field is the decisive parameter.
We study the mass spectrum of the DS, and outline regions of parameter space which satisfy constraints from the LHC searches on the invisible width of the
Higgs boson, and from Planck Space Telescope observations on the relic density of the CDM content.

\section{A Conformal Model for Dark Matter}
\label{sec:model}

A CDM candidate which belongs to a dark sector (DS) and is composed of SM singlets, can couple to the SM fields via Higgs,
hypercharge or neutrino portals. These interactions, invariant under both SM and DS gauge symmetries, already
exist at the renormalizable level, and exhibit conformal invariance if CDM particles are charged under a dark gauge
symmetry. Even when the DS is not governed by a gauge symmetry, as mentioned above, longevity of the CDM particle necessitates at least a
${\mathbb{Z}}_2$ symmetry. It is thus conceivable to consider a conformally-invariant DS which couples conformally to the SM fields. This conformal setup has the advantage that
a ${\mathbb{Z}}_2$ symmetry is inherently incorporated. 

Motivated by the discovery of the Higgs boson, which exhibits all the properties appropriate for an SM-like Higgs, and the wealth of 
experimental information supporting the SM, we adopt the SM as is, and impose that only the dark matter candidate 
obeys conformal invariance. Previous authors have investigated cases in which both the dark matter scalar and the 
Higgs boson are  conformally invariant \cite{Okada:2012sg}.  However, our aim here is to show that a minimally modified 
SM by the addition of a conformal dark matter candidate can satisfy bounds from both dark matter and invisible Higgs width. 
This scenario does not solve the fine-tuning problem for the additional scalar particle, though one can rely on alternative 
solutions, such as additional symmetries or particles to resolve it. For instance,  in \cite{Chakraborty:2012rb}  the fine-tuning 
problem of singlet scalar is resolved by adding $SU(2)$ singlet or doublet vector fermions such that the mass-squared value 
of the singlet scalar is protected against quantum corrections.

The main ingredient of our model is a conformally-invariant scalar field that couples conformally to the SM Higgs doublet.
The scalar field, an SM-singlet belonging to the DS, can be a real scalar $S$, or a complex scalar $\phi$, charged under a dark gauge group $U(1)_D$. This group contains a 
 gauge boson $A_{\mu}^{\prime}$, and, in addition, the DS sector can include a dark fermion $\psi$ charged under $U(1)_D$.\footnote{Higher-rank gauge groups do not bring any
further insight so we shall contend ourselves with a simple $U(1)_D$ symmetry.} Below, we investigate these fields one by one.

\subsection{Dark Real Scalar}
\label{subsec:realS}
The Higgs doublet $H$ and real singlet $S$ interact via
\bea\label{Lag1}
\mathcal{L}_{\textrm{S}}&=&(D_{\mu}H)^{\dagger}D^{\mu}H+\frac{1}{2}\,\partial_{\mu}S\,\partial^{\mu}S-V_S,
\eea
where the conformal-invariant potential energy
\bea
\label{pot1}
\!\!\!V_S\!\!=\!\frac{m_{H}^{2}}{2}H^{\dagger}H
\!+\frac{\lambda_H}{4}(H^{\dagger}H)^2\!+\frac{\lambda_{S}}{4}S^4\!-\frac{\lambda}{4}H^{\dagger}HS^2,
\eea
involves no interaction with scaling dimension different than $4$ ($S$, $S^2$, $S^3$, $S^5$ and so on), thus giving rise to automatically ${\mathbb{Z}}_2$-symmetric interactions for $S$. The only exception is $H$; its mass parameter $m_H^2$ generates all the scales in the  DS, and in the SM upon electroweak breaking.
With $\lambda_{H}>0$ and $\lambda_{S}>0$, the  potential is bounded from below, and its  minimization yields a phenomenologically interesting scenario where, for $m_{H}^{2}<0$, there is a
local maximum at $\langle 0|H|0\rangle \equiv \upsilon_H = 0$, $\langle0|S|0\rangle \equiv \upsilon_S = 0$, and a minimum at
\bea
\upsilon_H^2=-\frac{m_{H}^{2}}{\lambda_{H}}, \qquad  \upsilon_{S}^2=0.
\eea
For excitations of $H$ above the vacuum
\bea
\label{eq:higgs}
H=\frac{1}{\sqrt{2}}\left(\baa{c}
H_{3}+i H_{4}\\
\sqrt{2}\upsilon_H+H_{1}+i H_{2}
\eaa
\right),
\eea
we obtain a diagonal mass matrix for $H_1$ and $S$ (the massless $H_{2,3,4}$ are Goldstone bosons eaten by $W^{\pm}$ and $Z$). Here $H_1 \equiv h$
is the SM Higgs boson (with the additional interaction $\frac{\lambda}{4} H^{\dagger}H S^2$ in Eq. (\ref{pot1}) above). After electroweak  breaking
$S$ acquires mass, and conformal symmetry gets broken  to ${\mathbb{Z}}_2$ parity. The mass-squared of $S$ is proportional to $\lambda$
so that, as anticipated before, the smaller the $|\lambda|$, the lighter the real singlet scalar $S$. The model thus accommodates a naturally light, weakly interacting, stable scalar sector
which can set a standard for studies on light singlet scalar fields \cite{Silveira:1985rk}. The masses of the scalar fields
\bea
m_h^2=\lambda_H \upsilon_H^2, \qquad m_S^2=-\frac{\lambda}{2}\upsilon_H^2,
\eea
exhibit the hierarchy, $m_h^2 \gg m_S^2$, if $|\lambda|$ is small enough.

\subsection{Dark Complex  Scalar }
\label{subsec:complexS}

For the complex scalar $\phi$, interactions with Higgs field are encoded in
\bea\label{Lag2}
&&\mathcal{L}_\phi=(D_{\mu}H)^{\dagger}D^{\mu}H+\partial_{\mu}\phi^* \partial^{\mu}\phi-V_\phi,
\eea
where the conformal-invariant potential
\bea
\!\!\!\!\!\!\!\!\!\!V_\phi\!=\!\frac{m_{H}^{2}}{2}H^{\dagger}H\!+\!\frac{\lambda_H}{4}(H^{\dagger}H)^2\!
+\!\!\frac{\lambda_{\phi}}{4}(\phi^*\phi)^2\!-\!\frac{\lambda}{4}H^{\dagger}H\phi^*\phi,
\eea
has the same  structure as the potential in Eq. (\ref{pot1}). Thanks to conformal symmetry, it retains a ${\mathbb{Z}}_2$ symmetry associated with $\phi$.
The potential is bounded from below for $\lambda_{H}>0$ and $\lambda_{\phi}>0$, and possesses a phenomenologically interesting
minimum at
\bea
\upsilon_H^2=\frac{4\lambda_{\phi}m_{H}^{2}}{\lambda^2-4\lambda_{H}\lambda_{\phi}},\qquad
\upsilon_{\phi}^2=\frac{2\lambda m_{H}^{2}}{\lambda^2-4\lambda_{H}\lambda_{\phi}}.
\eea
%
Parametrizing $H$ as in (\ref{eq:higgs}) and the complex
scalar as $\phi=\frac{1}{\sqrt{2}}(\sqrt{2}\upsilon_{\phi}+\phi_1+i\phi_2)$, the $H_1$ and $\phi_1$ mix with through mass matrix
\bea
M^2_{H_1,\phi_1}=\left(
\baa{cc}
\lambda_H\upsilon_H^2 & -\frac{\lambda}{2}\upsilon_H\upsilon_{\phi}\\
-\frac{\lambda}{2}\upsilon_H\upsilon_{\phi} & \lambda_{\phi}\upsilon_{\phi}^2
\eaa
\right),
\eea
where now the Goldstone sector involves also $\phi_2$. After diagonalization,  this mass matrix yields the physical scalars $h$ and $\varphi$ with
masses
\bea
\!\!\!\!\!\!\!\!m^2_{h,\varphi}\!=\!\frac{1}{4} (2\lambda_H +\lambda) \upsilon_H^2 \bigg(\!1 \pm\! \sqrt{1+ \frac{2 \lambda ( \lambda^2 - 4 \lambda_H \lambda_{\phi})}{\lambda_{\phi}
( 2 \lambda_H + \lambda)^2}}\bigg),
\eea
and mixing angle
\bea
\tan^2 2\theta=\frac{\lambda^{3}}{2\lambda_{\phi} (2\lambda_{H}-\lambda)^2}.
\eea
The complex field $\phi$ behaves differently than the real scalar $S$. The mass of $\varphi$ breaks conformal symmetry, the VEV of $\phi$ breaks ${\mathbb{Z}}_2$
parity, and hence, $\varphi$ cannot be a CDM candidate. A viable CDM candidate is found either in the gauge boson
of gauge $U(1)_D$ or in a dark fermion sterile under SM but charged under $U(1)_D$, which can be added to the spectrum. These candidates are analyzed below. The singlet scalars as a solution to fine-tuning problem have been discussed in \cite{Kundu:1994bs}. It has been shown in \cite{Farzinnia:2013pga} that, with no extra global or local symmetry introduced, the imaginary component of the complex singlet scalar $\phi_2$ is also a viable DM candidate, and that its stability is automatically protected by CP invariance.

\subsection{Vacuum Stability Conditions}

The tree-level potential minimum is simply guaranteed by requiring
\begin{equation}
\lambda^2-4\lambda_{H}\lambda_{\phi}>0, 
\end{equation}
while the  requirement that the potential is bounded from below is:
\begin{equation}
\lambda_{H}>0~~~{\rm and}~\lambda_{\phi}>0.
\end{equation}
A full two-loop analysis of the vacuum stability conditions  would lead to a precise statement of perturbativity for the quartic couplings and a more restricted range of
parameter space, but this is beyond the scope of this work.  We rely on previous analyses \cite{Hambye:1996wb}, which  considered two possible criteria to
constrain the values of the couplings at the cut-off scale, thus leading to a perturbative loop expansion of the potential. The first option is to take the SM
two-loop result and apply it to each of the quartic couplings at the cut-off scale individually. The second, less restrictive, is to follow the constraints of \cite{Lerner:2009xg} 
\begin{eqnarray}
\!\!\!\!\!\!\!\!\!\!\!\!\!\!\!\!\!\!\!\!\!\!\!&& \lambda_H<8\pi/3,\, \lambda_{S}<2\pi/3,\,\lambda<8\pi\,\,\,\,  {\rm for\,real\,singlet, } \nonumber\\
\!\!\!\!\!\!\!\!\!\!\!\!\!\!\!\!\!\!\!\!\!\!\!&&\lambda_H<8\pi/3,\, \lambda_{\phi}<8\pi/3,\, \lambda<16\pi\,\,\,\,  {\rm for\,complex\,singlet},\nonumber 
\end{eqnarray}
where the perturbativity condition $\lambda_i'<4\pi$ is used.
We choose this scenario, as the constraints are cut-off independent. In Section \ref{sec:pheno}, we shall see that these conditions are comfortably satisfied by our parameters in the region of phenomenological interest, as all the couplings in our model are potentially less than $0.1$ to satisfy the relic density bounds.

\subsection{Dark Gauge Boson}
Gauged $U(1)_D$ modifies the original Lagrangian (\ref{Lag2}) by contributions to the kinetic term via
%
\label{covariant}
$\partial_{\mu}\phi\rightarrow D_{\mu}\phi=(\partial_{\mu}-ie_{_{D}} A_{\mu}')\phi$,
%
where $e_D$ is the $U(1)_D$ gauge coupling. The Lagrangian
\bea 
{\cal L}_{A'}= {\cal L}_\phi - \frac{1}{4}F_{\mu\nu}'F'^{\mu\nu}
\eea
where ${\cal L}_\phi$ is the complex scalar Lagrangian from Eq. (\ref{Lag2}). The
$A_{\mu}^{\prime}$ acquires the mass
\bea
m_{A'}^2=\frac{\lambda}{\lambda_{\phi}} e_{_D}^2\upsilon_H^2,
\eea
from $U(1)_D$ breaking. Possible kinetic mixing between $U(1)_D$ and hypercharge $U(1)$ are avoided by imposing
$A_{\mu}'\rightarrow -A_{\mu}'$ and $\phi\rightarrow \,\phi*$ invariance \cite{Farzan:2012hh}. The gauged vector CDM
models have been studied in \cite{Farzan:2012hh, Servant:2002aq}.

\subsection{Dark Fermion}
Just like the scalars $S$ or $\phi$, there can exist a dark fermion $\psi$ in DS. It can be the CDM by itself or in
addition to the $A_{\mu}^{\prime}$ and the real scalars $S$. As a sterile fermion charged under $U(1)_D$,
it can interact only with $\phi$
\bea
\label{majorana}
\mathcal{L}_{\psi}=\bar{\psi} i\not\!\!{D} \psi + \frac{\lambda_{\psi}}{2}\phi\,\bar{\psi}^c\psi,
\eea
where $\psi^c$ is charge-conjugate of $\psi$, and $U(1)_D$ charges satisfy $Q_{\phi}=2 Q_{\psi}$. Upon $U(1)_D$ breaking, the dark fermion
acquires the mass
 \bea
 m_{\psi}^2=\lambda_{\psi}\upsilon_{\phi}^2=\frac{\lambda_{\psi}\lambda}{2\lambda_{\phi}}\upsilon_H^2,
 \eea
which is proportional to $\lambda$. This fermion accesses the SM fields via $h$--$\varphi$ mixing. (See \cite{Dodelson:1993je} for a similar models
with sterile neutrinos.)
\section{Phenomenological Implications}
\label{sec:pheno}
The DS fields studied above can have important impact on collider experiments and astrophysical observations.
While a detailed analysis can shed more light on the model parameters, we here focus exclusively on
the Higgs invisible rate measured at the LHC and the current CDM density reported by PLANCK 2013, and show that the experimental data can be satisfied within each scenario, and for a minimal number of parameters.

\subsection{Bounds from Higgs Invisible Width}
If kinematically allowed, the Higgs can decay into the dark matter candidates. Then the decay $h\rightarrow XX$,  $X=S, \varphi, \psi, A'_{\mu}$ constitute the Higgs invisible rate $\Gamma_{\textrm{inv}}$, and is constrained by measurements of the Higgs width. For a given $X$ the width is given by
\bea
\Gamma_{h\rightarrow XX} = \frac{|\mathcal{M}(X)|^2}{32\,\pi\,m_{X}}\sqrt{1-4\frac{m_{X}^2}{m_h^2}},
\eea
where the matrix elements for real scalars, vector bosons and complex scalars are, respectively,
\bea
|\mathcal{M}(S)|^2\,&=&\,\frac{\lambda^2}{8}\upsilon_H^2,\qquad |\mathcal{M}(\psi)|^2=\frac{\lambda_{\psi}^2}{4}\sin^2\theta\, m_h^2,\non\\
|\mathcal{M}(A')|^2&=&\frac{\lambda\upsilon_{H}^2e_D^4\sin^2\theta}{4\lambda_{\phi}m_{A'}^4}(12m_{A'}^4-4m_{A'}^2m_h^2+m_h^4),\non\\
|\mathcal{M}(\phi)|^2&=&\bigg|\frac{3\upsilon_H\sin2\theta}{4\sqrt{2}}\!\left(\lambda_H\sin\theta-\sqrt{\frac{\lambda\lambda_{\phi}}{2}}\cos\theta\right)\non\\
&&-\frac{\lambda\upsilon_H}{4\sqrt{2}}f(\theta)+\sqrt{\frac{\lambda}{2\lambda_{\phi}}}g(\theta)\bigg|^2.
\eea
Here $f(\theta)=\cos^3\theta-\sin2\theta\sin\theta$ and $g(\theta)=\sin^3\theta-\sin2\theta\cos\theta$.
For $m_h=126$ GeV,
$\Gamma^{\textrm{vis}}_=4.21$ MeV \cite{Dittmaier:2012vm},  and Higgs invisible branching $\textrm{BR}^{\textrm{inv}}_h=\Gamma^{\textrm{inv}}_h/(\Gamma^{\textrm{inv}}_h+
\Gamma^{\textrm{vis}}_h)$ is constrained to be less than $19\%$ at $2\sigma$ \cite{Belanger:2013xza}.

\begin{figure}[h]
 \subfigure[] {\includegraphics[width=0.45\textwidth]{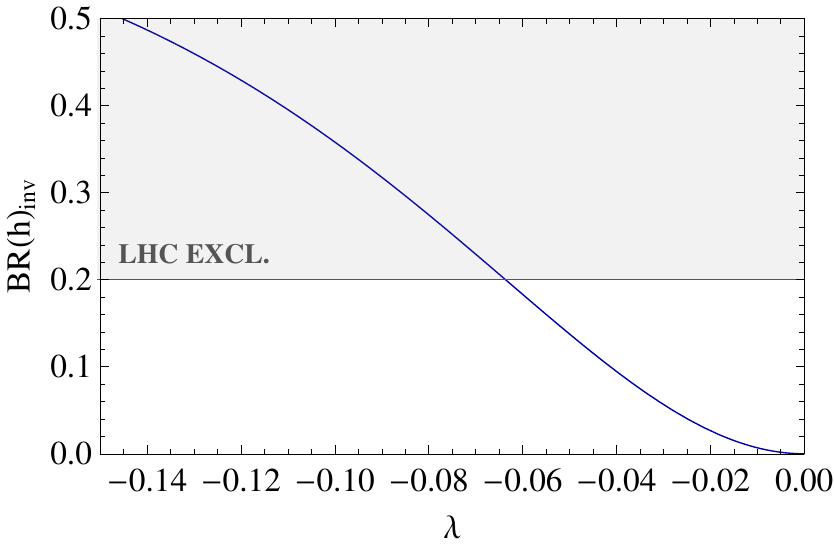}}
\caption{$\textrm{BR}^{\textrm{inv}}_h$ for $X=S$ (blue)  with LHC exclusion region (in light grey).}
\label{fig:BRRSinv}
\end{figure}

If DS involves just $S$, $\Gamma^{\textrm{inv}}_h = \Gamma_{h\rightarrow SS}$, and the invisible width depends only $\lambda$. In
Fig. \ref{fig:BRRSinv} we plot  the bound on $\lambda$ from $\textrm{BR}^{\textrm{inv}}_h$. The LHC-allowed region (represented by the white part of the plot) corresponds to
$\lambda \gtrsim- 0.063$  (yielding $m_{S}\lesssim 30$ GeV for the DM real scalar).

\begin{figure}[h]
\begin{center}
\vskip-0.5in
 \subfigure[] {\includegraphics[width=0.35\textwidth]{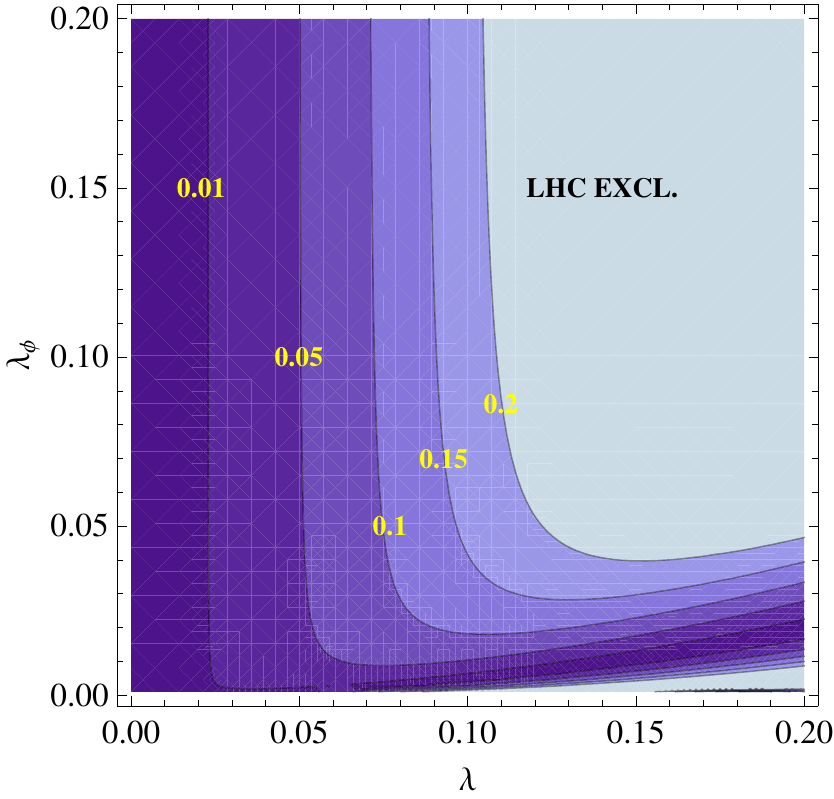}}
  \subfigure[] { \includegraphics[width=0.35\textwidth]{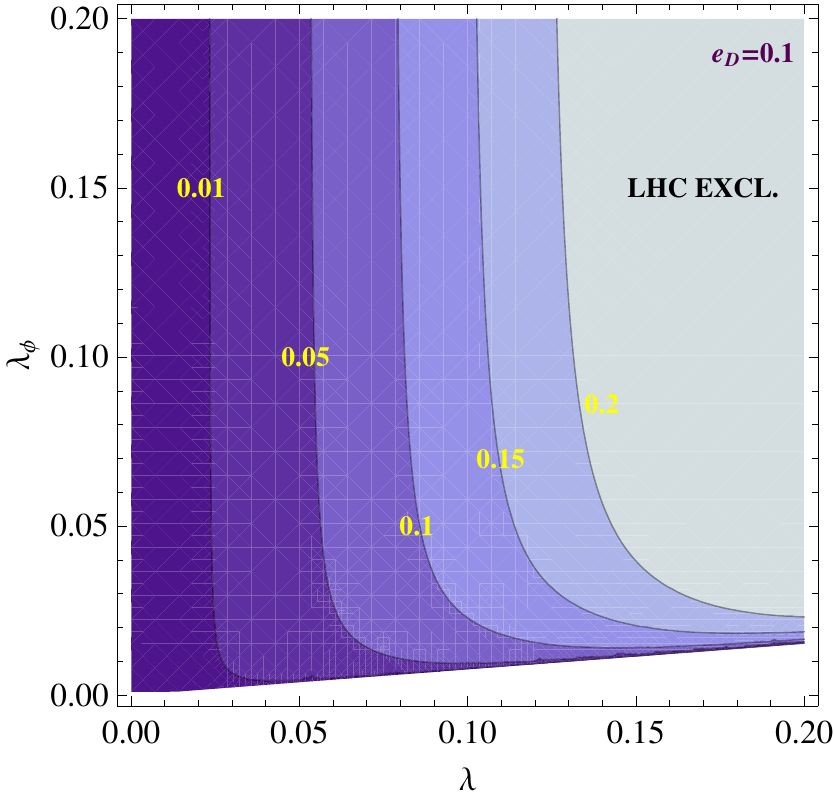}}
  \subfigure[] {\includegraphics[width=0.35\textwidth]{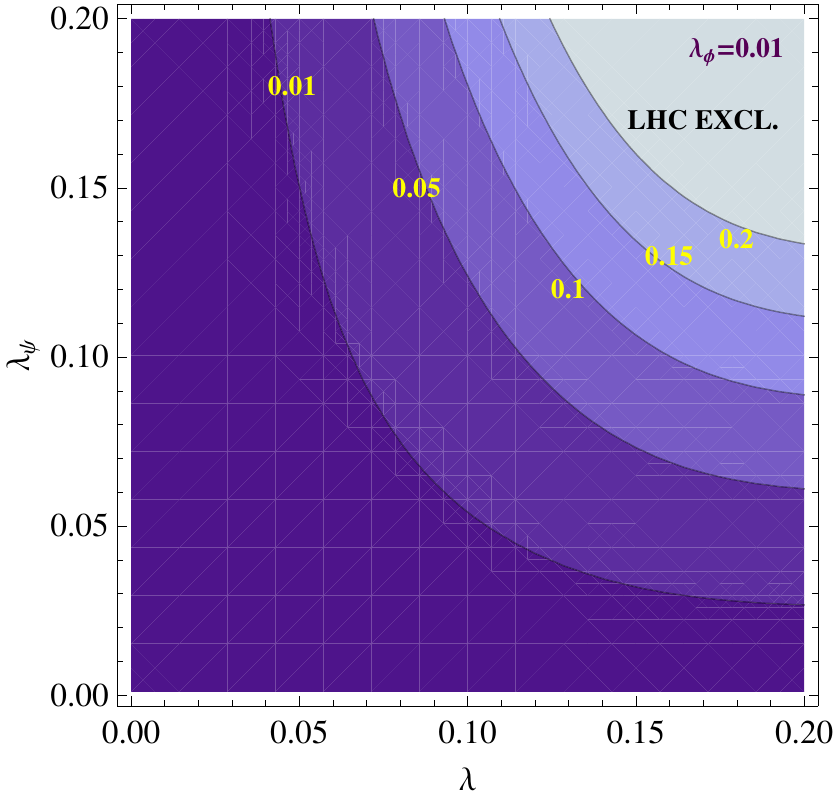}}
\vskip -0.1in
\caption{Contour plots of $\textrm{BR}^{\textrm{inv}}_h$ for $X=\varphi$ (top), $X=A_{\mu}^{\prime}$ (middle), and
$X=\psi$ (bottom) with LHC exclusions (light-blue).}
\label{fig:BRCSinv}
\end{center}
\end{figure}

If DS consists of the complex scalar $\phi$ then $\Gamma^{\textrm{inv}}_h$ can be either $\Gamma_{h\rightarrow \varphi\varphi}$ or
$\Gamma_{h\rightarrow A' A'}$ or $\Gamma_{h\rightarrow \psi\psi}$. They each are plotted in Fig. \ref{fig:BRCSinv}
for different $\textrm{BR}^{\textrm{inv}}_h$ values. In general, LHC-excluded regions occur at larger
values of couplings, and are depend on other couplings. For instance, if $e_D$ is increased,
the light grey region moves to upper right corner with $\lambda$ and $\lambda_{\phi}$ getting closer to $e_D$
($\lambda_{\phi} \gtrsim 0.17, \lambda\gtrsim 0.2$).
 
\subsection{Bounds from Relic Density}
The relic density of CDM today is determined by rate at which the DM  coannihilate into SM  states ($X X \rightarrow\, SM$).
The rate is dominated by $s$-channel annihilation through mediator $X=h$ or $\varphi$  using a  Breit-Wigner propagator
\bea
D_X(s)=\frac{1}{s-m_X^2+i m_X\,\Gamma_{X}^{\textrm{tot}}},
\eea
where $\Gamma_{X}^{\textrm{tot}}$ is the total width (visible plus invisible) of $X$.

\begin{center}
\begin{figure}[h]
\vskip-1.5in
  \subfigure[] {\includegraphics[width=0.4\textwidth]{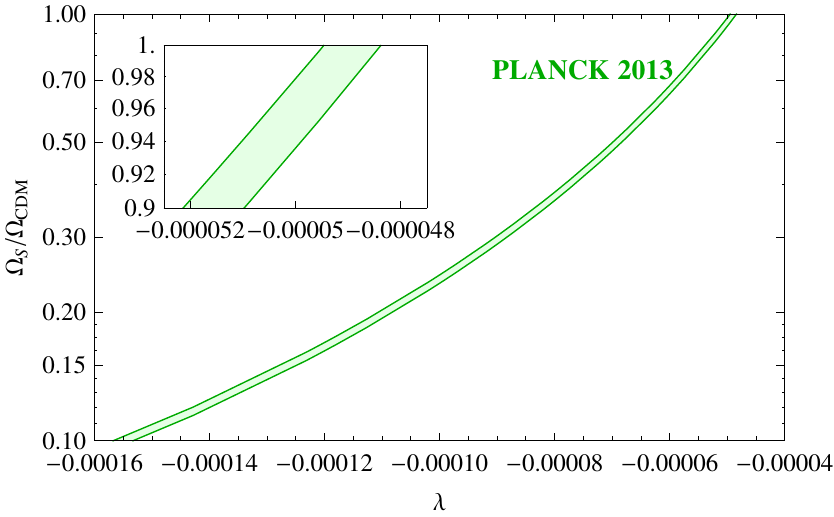}}
  \subfigure[] {\includegraphics[width=0.35\textwidth]{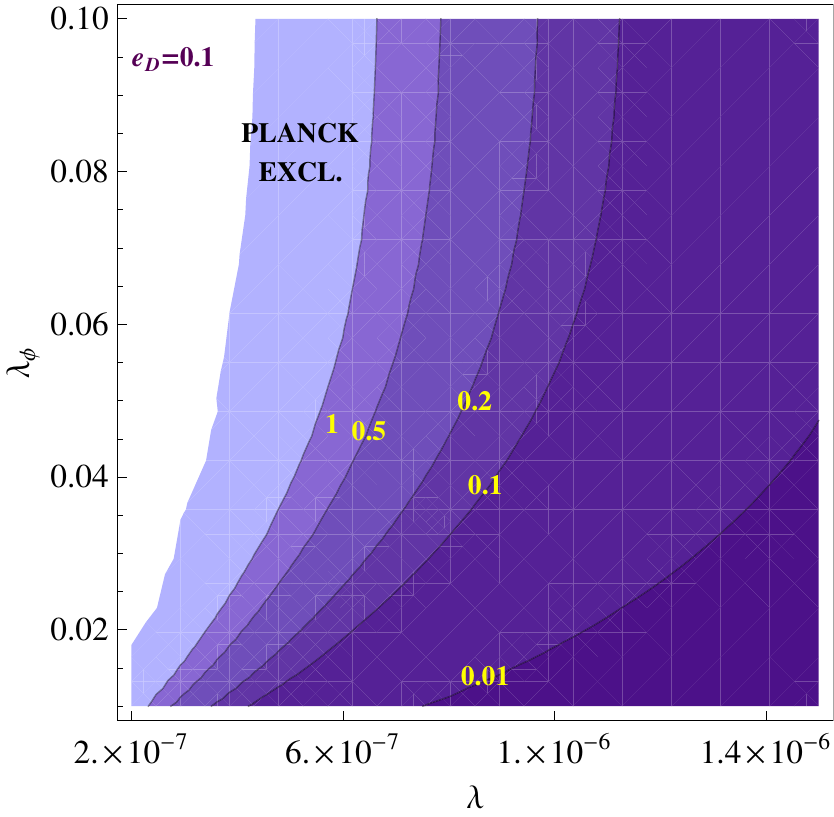}}
   \subfigure[] {\includegraphics[width=0.35\textwidth]{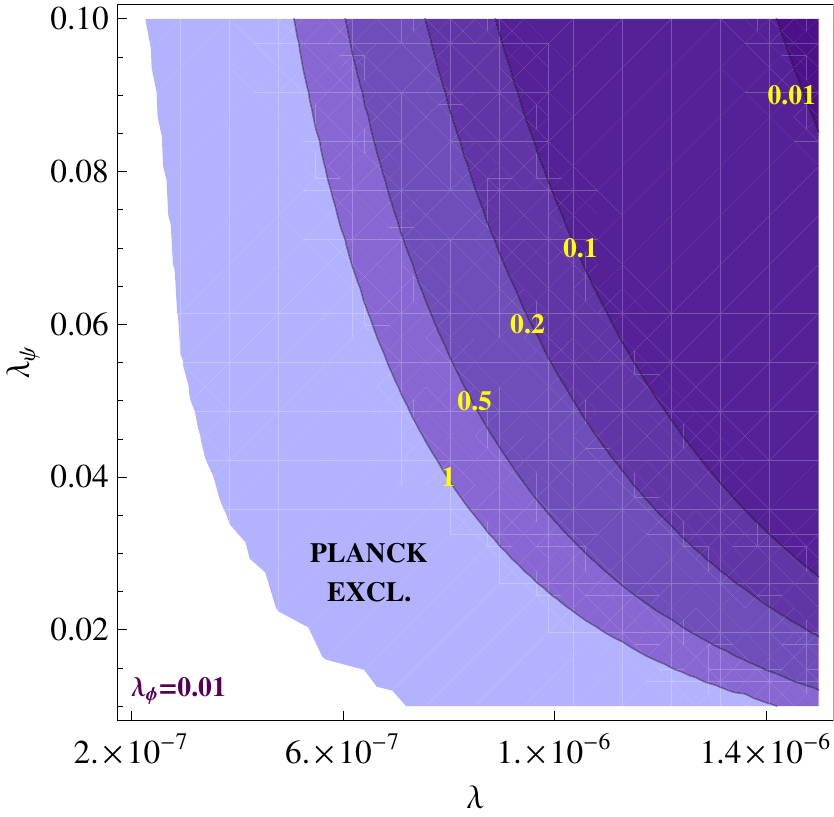}}
\vskip -0.1in
\caption{Relic abundance plots of the CDM candidates $S$ (top), $A_{\mu}^{\prime}$ (middle) and $\psi$ (bottom). Contours indicate the ratio $\Omega_X/\Omega_{\rm CDM}$.}
\label{fig:relics}
\end{figure}
\end{center}

For $X=S$, the relic density
\bea
\langle\sigma v_{\textrm{rel}}\rangle=\frac{\lambda^2\upsilon_H^2}{4\sqrt{s}}\,|\,D_h(s)|^2\,
\,\Gamma_h^{\textrm{vis}}(\sqrt{s}),
\eea
is enhanced near the Higgs resonance, $\sqrt{s} \approx m_h$ at which  $\Gamma_h^{\textrm{vis}}(\sqrt{s})$ becomes the visible width of the Higgs boson
\cite{Dittmaier:2012vm}.  In the left panel of Fig \ref{fig:relics} we depict the ratio $\Omega_S/\Omega_{\textrm{CDM}}$, where $\Omega_S$ is the relic density of $S$, as a function of $\lambda$. The 
2013 PLANCK measurement of the relic density is $\Omega_{\rm CDM} h^2=0.1199\pm0.0027$ at $68\%$ CL \cite{Ade:2013ktc}, and is represented in the figure by the thin green strip. The figure shows that
a single real scalar singlet is able to saturate entire CDM density if $-4.8\times 10^{-5}\gtrsim\lambda\gtrsim -4.9\times 10^{-5}$
(or $0.73 \textrm{GeV}\lesssim m_S \lesssim 0.75 \textrm{GeV}$). This narrow region is consistent with $\lambda \gtrsim -0.063$ determined
from Fig. \ref{fig:BRRSinv}.

In the case of the SM particle content being augmented by a complex scalar $\phi$, the dark gauge boson $A_{\mu}^{\prime}$ and/or the dark fermion $\psi$ can serve as CDM particles.
The relic density of $A_{\mu}^{\prime}$, diluting through $A'_{\mu}A'_{\mu}\rightarrow \textrm{SM}$ annihilation, is given by
\bea
\!\!\!\!\langle\sigma v_{\textrm{rel}}\rangle\!\!\!\!\!&=&\!\!\!\!\!\frac{e_D^4\lambda\upsilon_H^2}{2\lambda_{\phi}\sqrt{s}m_{A'}^4}
|D(s)|^2\big(12m_{A'}^4-4m_{A'}^2s+s^2\big),\;\;\;
\eea
through $h$ and $\varphi$ mediators encoded in the propagator
\bea
\!\!\!\!\!\!\!\!\!\!D(s)\!\!\!\!\!&=&\!\!\!\!\!D_{h}(s)\sin\theta
\sqrt{\frac{\Gamma_h(s)\Gamma_h^{\textrm{vis}}}{\Gamma_h(m_h)}}-D_{\varphi}(s)\cos\theta
\sqrt{\frac{\Gamma_{\varphi}(s)\Gamma_{\varphi}^{\textrm{vis}}}{\Gamma_{\varphi}(m_{\varphi})}}.\non
\eea
Here $\Gamma_h$ is the Higgs fermionic width
\bea
 \Gamma_h(s)&=&\frac{N_c^f m_f^2\sqrt{s}}{16\pi\upsilon_H^2}\left(1-4\frac{m_f^2}{s}\right)^{3/2}\cos^2\theta,\non
 \eea
and   $\Gamma_{\varphi}(s)=  \Gamma_h(s) (\cos \theta \to \sin\theta)$ is the $\varphi$ width.  The relic density of $A'_{\mu}$ is enhanced at $\sqrt{s}=m_{\varphi}$, and 
almost independent of $e_D$. The ratio $\Omega_{A'}/\Omega_{\rm CDM}$ is plotted in the middle panel of Fig. \ref{fig:relics} for $e_D=0.1$. This plot is consistent with Fig. \ref{fig:BRRSinv}. Clearly,
$A_{\mu}^{\prime}$, now a viable CDM candidate, can saturate the CDM in the Universe either all by itself,  or partially.

The relic density of the dark fermion
\bea
\langle\sigma v_{\textrm{rel}}\rangle&=&\frac{\lambda_{\psi}^2\sqrt{s}}{2}|D(s)|^2.
\eea
is also pronounced at $\sqrt{s}\sim m_{\varphi}$. From the right panel of Fig. \ref{fig:relics}, where we plot the relic density due to the dark fermion only, it is clear that this fermion is a viable CDM candidate. 
The panels of this figure, separately or together, can account for the abundance of CDM in the Universe for wide ranges of parameters.

\subsection{Dark Matter Searches and Astrophysical Constraints}

Our model predicts candidates for dark matter of low mass, which have, up  to now,  shown resilience to experimental constraints.
Indirect constraints come from the Cosmic Microwave Background, gamma rays and neutrino experiments in particular. Several observations of cosmic and gamma-ray 
fluxes have been linked to the possible signals
of annihilation or decays of DM particles. The  511 keV line emission from the Galaxy detected by the SPI spectrometer on the INTEGRAL satellite \cite{Knodlseder:2003sv}, the excesses of microwaves and gamma rays in the inner Galaxy revealed by the WMAP and Fermi satellites \cite{Su:2010qj}, the evidence for a 130 GeV spectral line in the Fermi data \cite{Bringmann:2012vr}, which predicts that DM particles with masses below 10 GeV  have the annihilation cross section at $\sigma_{\rm th}\upsilon\sim3\times 10^{-26} {\rm cm}^3/{\rm s} $, or the rise in the positron fraction above 10 GeV observed by PAMELA \cite{Adriani:2008zr} and AMS-02 \cite{AMS}, which even though it could be interpreted as the signal from nearby pulsars or astrophysical objects \cite{Hooper:2008kg}, it still provides stringent bounds on the DM annihilation cross section to electron-positron or muon-antimuon
pairs \cite{Bergstrom:2013jra}, and the current constraint on dark matter scattering with nuclei largely through spin-dependent couplings from the IceCube experiment \cite{Aartsen:2012kia}, all have been interpreted as physics associated with DM \cite{indirectDM}. Currently a
dark matter interpretation for these signals is far from clear, given limited statistics (for the Fermi line) or large systematics or astrophysical backgrounds (for the positron, and the 511 keV emission), as shown in an up-to-date review of indirect searches, see \cite{Buckley:2013bha}.

In direct searches, various anomalies have remained, 
while new constraints have continued to close the allowed parameter space for an elastically scattering light DM particle in the 7-12 GeV mass region that can explain the signals.

These anomalies have become the target for searches of light DM, and the null results from
XENON10 \cite{Angle:2011th}, XENON100 \cite{Aprile:2012nq}, PICASSO \cite{Archambault:2012pm}, COUPP \cite{Behnke:2012ys}, CDMS-Ge low energy \cite{Ahmed:2009zw} and CDMSlite  \cite{Agnese:2013lua} constrained the region. The strongest constraints are obtained from XENON in the spin-independent case,  but they are subject to nuclear recoil energy calibration uncertainties near  the threshold \cite{Aprile:2012nq}. 

CDMS-Si \cite{Agnese:2013rvf} reported an excess of three events at threshold consistent with a light DM candidate.  The preferred region appears consistent with the excess observed by CoGeNT \cite{Aalseth:2012if}, though may be in conflict with the XENON100 constraint.\footnote{Since the targets in CoGeNT \cite{Aalseth:2012if}, DAMA \cite{Bernabei:2013cfa}, and CDMS are different than in XENON100, the constraints are difficult to interpret in a model-independent scenario \cite{Gresham:2013mua}.}

Most recently, LUX has published data on light DM with a low nuclear recoil energy
 threshold of 3 keV \cite{Akerib:2013tjd}. For the three CDMS events, assuming equal DM coupling to the proton and neutron, and spin-independent scattering, at a cross-section $2 \times10^{-41}$ cm$^2$, LUX would be expected to see approximately 1500 events. Thus LUX is able to put a strong constraint on the entire preferred region of the CDMS-Si three events.

The results from LUX  rule out the region where all three experiments overlap.  PICASSO, XENON10  and CDMS-Ge low-energy are also competitive in this range, and various dark matter models offer alternative assumptions to weaken the LUX constraint relative to the CDMS-Si and CoGeNT regions of interest. However, when corrected for energy nuclear recoil energy calibration, LUX provides the strongest bounds on dark matter masses above 5.5 GeV. Neither LUX or XENON100 are sensitive below this threshold \cite{DelNobile:2013gba}, and our results still stand.

In addition, observations at Bullet Cluster \cite{Clowe:2006eq} could be used to place a constraint on the quartic DM coupling $\lambda_S$, as the ratio DM scattering cross section over the mass must be less than 1.25 cm$^2/$g, would imply a lower bound on the mass of the dark matter candidate,  $m_{\rm DM}>64$ MeV \cite{Bento:2000ah}, consistent with what we have obtained here. 
See also \cite{Arrenberg:2013rzp} and references therein for a comprehensive review on direct and indirect DM searches.

\clearpage
\section{Conclusion}
\label{sec:conclusion}
In this work, we constructed a CDM model by augmenting the SM particle content by  a conformal-invariant dark sector, interacting conformally with
the SM through the Higgs portal, and including either a singlet real, or a complex scalar field. While the real singlet does not develop a VEV and can 
become itself a DM candidate, the complex scalar scenario has to be augmented by a dark sector containing a vector gauge boson, or a fermion charged 
under an additional symmetry $U(1)_D$.   The near conformal invariance of the SM, with initially massless fermions and gauge bosons, supports the idea 
that, in beyond the SM scenarios, additional sectors should also respect the conformal symmetry. The only term which breaks the conformal symmetry 
is the squared mass of the Higgs, which after electroweak symmetry breaking, is responsible for the scales for both the SM and DS. Our main motivation 
for introducing conformal invariance is that it naturally embodies a ${\mathbb{Z}}_2$ symmetry, which
is crucial for stability of DM.  As a result of conformal invariance, the lighter the dark
sector fields, the weaker their couplings to Higgs boson, and this implies that dark matter and invisible Higgs decays are closely 
related phenomena, and that conformal invariance provides a natural framework that connects them.  We have shown that either the  scalar singlet, or the gauge 
boson and fermion, alone or in combination, respect constraints on the invisible width of the Higgs boson, as inferred at LHC,  and comfortably satisfy 
measurements from PLANCK 2013 on relic density, while being consistent with the  constraints from dark matter searches. Thus the model presented here  provides viable scenario for cold dark matter. Though additional particles and interactions would need to be added to resolve some other outstanding issues in the SM, such as the hierarchy problem, the strength of the model 
lies in its simplicity and minimal number of parameters, thus predictability.  This model can be an integral part of the Project X efforts \cite{projectX}.

\section{Acknowledgements}
We thank T\"{U}B\.{I}TAK, The Scientific and Technical Research Council of Turkey, through the grant 2232, Project No: 113C002, and  NSERC of Canada (under grant  SAP105354) for financial support. DAD and BK thank Yasaman Farzan for fruitful discussions.



\begin{thebibliography}{99}

\bibitem{Aad:2012tfa}
  G.~Aad {\it et al.}  [ATLAS Collaboration],
  Phys.\ Lett.\ B {\bf 716} (2012) 1;
  S.~Chatrchyan {\it et al.}  [CMS Collaboration],
  Phys.\ Lett.\ B {\bf 716}, 30 (2012).

\bibitem{Ade:2013uln} 
  P.~A.~R.~Ade {\it et al.}  [Planck Collaboration],
  arXiv:1303.5082 [astro-ph.CO].

\bibitem{Jungman:1995df}
  G.~Jungman {\it et al.},
  Phys.\ Rept.\  {\bf 267} (1996) 195.

\bibitem{Hooper:2007qk}
  D.~Hooper and S.~Profumo,
  Phys.\ Rept.\  {\bf 453} (2007) 29.
  
\bibitem{Susskind:1978ms}
  L.~Susskind,
  Phys.\ Rev.\ D {\bf 20} (1979) 2619.

\bibitem{Bardeen:1995kv}
  W.~A.~Bardeen,
  FERMILAB-CONF-95-391-T.
  
\bibitem{Hur:2007uz}
  T.~Hur, D.~-W.~Jung, P.~Ko and J.~Y.~Lee,
  Phys.\ Lett.\ B {\bf 696} (2011) 262;
  T.~Hur and P.~Ko,
  Phys.\ Rev.\ Lett.\  {\bf 106} (2011) 141802.

\bibitem{Demir:2012nd}
  D.~A.~Demir,
  arXiv:1207.4584 [hep-ph].

\bibitem{Farzinnia:2013pga}
A.~Farzinnia, H.~-J.~He and J.~Ren,
  arXiv:1308.0295 [hep-ph].
  
\bibitem{Okada:2012sg} 
  N.~Okada and Y.~Orikasa,
  Phys.\ Rev.\ D {\bf 85}, 115006 (2012);
  P.~H.~Frampton,
  Mod.\ Phys.\ Lett.\ A {\bf 22}, 931 (2007).
\bibitem{Chakraborty:2012rb}
  I.~Chakraborty and A.~Kundu,
  Phys.\ Rev.\ D {\bf 87} (2013) 055015.
  

  

\bibitem{Silveira:1985rk} 
  V.~Silveira and A.~Zee,
  Phys.\ Lett.\ B {\bf 161}, 136 (1985);
  J.~McDonald,
  Phys.\ Rev.\ D {\bf 50}, 3637 (1994);
  C.~P.~Burgess, M.~Pospelov and T.~ter Veldhuis,
  Nucl.\ Phys.\ B {\bf 619}, 709 (2001);
  H.~Davoudiasl, R.~Kitano, T.~Li and H.~Murayama,
  Phys.\ Lett.\ B {\bf 609}, 117 (2005);
  S.~W.~Ham, Y.~S.~Jeong and S.~K.~Oh,
  J.\ Phys.\ G G {\bf 31}, 857 (2005);
  B.~Patt and F.~Wilczek,
  [hep-ph/0605188];
  D.~O'Connell, M.~J.~Ramsey-Musolf and M.~B.~Wise,
  Phys.\ Rev.\ D {\bf 75}, 037701 (2007);
  X.~-G.~He, T.~Li, X.~-Q.~Li and H.~-C.~Tsai,
  Mod.\ Phys.\ Lett.\ A {\bf 22}, 2121 (2007);
  S.~Profumo, M.~J.~Ramsey-Musolf and G.~Shaughnessy,
  JHEP {\bf 0708}, 010 (2007);
  V.~Barger, P.~Langacker, M.~McCaskey, M.~J.~Ramsey-Musolf and G.~Shaughnessy,
  Phys.\ Rev.\ D {\bf 77}, 035005 (2008);
  X.~-G.~He, T.~Li, X.~-Q.~Li, J.~Tandean and H.~-C.~Tsai,
  Phys.\ Rev.\ D {\bf 79}, 023521 (2009);
  E.~Ponton and L.~Randall,
  JHEP {\bf 0904}, 080 (2009);
  R.~N.~Lerner and J.~McDonald,
  Phys.\ Rev.\ D {\bf 80}, 123507 (2009);
  M.~Farina, D.~Pappadopulo and A.~Strumia,
  Phys.\ Lett.\ B {\bf 688}, 329 (2010);
  J.~M.~Cline, K.~Kainulainen, P.~Scott and C.~Weniger,
  Phys.\ Rev.\ D {\bf 88}, 055025 (2013);
%
  V.~Barger, P.~Langacker, M.~McCaskey, M.~Ramsey-Musolf and G.~Shaughnessy,
  Phys.\ Rev.\ D {\bf 79} (2009) 015018;
%
  M.~Gonderinger, H.~Lim and M.~J.~Ramsey-Musolf,
  Phys.\ Rev.\ D {\bf 86} (2012) 043511;
%
\bibitem{Kundu:1994bs}
  A.~Kundu and S.~Raychaudhuri,
  Phys.\ Rev.\ D {\bf 53} (1996) 4042;
  B.~Grzadkowski and J.~Wudka,
  Phys.\ Rev.\ Lett.\  {\bf 103} (2009) 091802;
  A.~Drozd {\it et. al},
  JHEP {\bf 1204} (2012) 006; 
  J.~R.~Espinosa and M.~Quiros,
  Phys.\ Rev.\ D {\bf 76} (2007) 076004;
  I.~Chakraborty and A.~Kundu,
  Phys.\ Rev.\ D {\bf 87} (2013) 055015. 


\bibitem{Hambye:1996wb} 
  T.~Hambye and K.~Riesselmann,
  Phys.\ Rev.\ D {\bf 55}, 7255 (1997); 
  K.~Riesselmann and S.~Willenbrock,
  Phys.\ Rev.\ D {\bf 55}, 311 (1997);
  M.~Gonderinger, Y.~Li, H.~Patel and M.~J.~Ramsey-Musolf,
  JHEP {\bf 1001}, 053 (2010).
  
\bibitem{Lerner:2009xg} 
  R.~N.~Lerner and J.~McDonald,
  Phys.\ Rev.\ D {\bf 80}, 123507 (2009).



\bibitem{Farzan:2012hh}
  Y.~Farzan and A.~R.~Akbarieh,
  JCAP {\bf 1210}, 026 (2012)
  [arXiv:1207.4272 [hep-ph]];
  Y.~Farzan and A.~R.~Akbarieh,
  arXiv:1211.4685 [hep-ph].

\bibitem{Servant:2002aq}
  G.~Servant and T.~M.~P.~Tait,
  Nucl.\ Phys.\ B {\bf 650} (2003) 391;
  A.~Birkedal, A.~Noble, M.~Perelstein and A.~Spray,
  Phys.\ Rev.\ D {\bf 74} (2006) 035002;
  T.~Hambye and M.~H.~G.~Tytgat,
  Phys.\ Lett.\ B {\bf 683}, 39 (2010);
  T.~Hambye,
  JHEP {\bf 0901}, 028 (2009);
  C.~Arina, T.~Hambye, A.~Ibarra and C.~Weniger,
  JCAP {\bf 1003}, 024 (2010);
  J.~L.~Diaz-Cruz and E.~Ma,
  Phys.\ Lett.\ B {\bf 695}, 264 (2011);
  J.~Hisano, K.~Ishiwata, N.~Nagata and M.~Yamanaka,
  Prog.\ Theor.\ Phys.\  {\bf 126}, 435 (2011);
  O.~Lebedev, H.~M.~Lee and Y.~Mambrini,
  Phys.\ Lett.\ B {\bf 707}, 570 (2012).


\bibitem{Dodelson:1993je}
  S.~Dodelson and L.~M.~Widrow,
  Phys.\ Rev.\ Lett.\  {\bf 72} (1994) 17;
  X.~-D.~Shi and G.~M.~Fuller,
  Phys.\ Rev.\ Lett.\  {\bf 82} (1999) 2832;
  K.~Abazajian, G.~M.~Fuller and M.~Patel,
  Phys.\ Rev.\ D {\bf 64} (2001) 023501;
  K.~Abazajian,
  Phys.\ Rev.\ D {\bf 73} (2006) 063506;
  M.~Shaposhnikov and I.~Tkachev,
  Phys.\ Lett.\ B {\bf 639} (2006) 414;
  Y.~G.~Kim and K.~Y.~Lee,
  Phys.\ Rev.\ D {\bf 75}, 115012 (2007);
  A.~Boyarsky, A.~Neronov, O.~Ruchayskiy and M.~Shaposhnikov,
  Phys.\ Rev.\ D {\bf 74}, 103506 (2006);
  Y.~G.~Kim and K.~Y.~Lee,
  Phys.\ Rev.\ D {\bf 75} (2007) 115012;
  K.~Petraki and A.~Kusenko,
  Phys.\ Rev.\ D {\bf 77} (2008) 065014;
  Y.~G.~Kim, K.~Y.~Lee and S.~Shin,
  JHEP {\bf 0805}, 100 (2008);
 S.~Kanemura, S.~Matsumoto, T.~Nabeshima and N.~Okada,
  Phys.\ Rev.\ D {\bf 82}, 055026 (2010);
  M.~Pospelov and A.~Ritz,
  Phys.\ Rev.\ D {\bf 84}, 113001 (2011);
  A.~Djouadi, O.~Lebedev, Y.~Mambrini and J.~Quevillon,
  Phys.\ Lett.\ B {\bf 709}, 65 (2012);
 S.~Baek, P.~Ko and W.~-I.~Park,
  JHEP {\bf 1202}, 047 (2012);
L.~Lopez-Honorez, T.~Schwetz and J.~Zupan,
  Phys.\ Lett.\ B {\bf 716}, 179 (2012);
S.~Baek, P.~Ko, W.~-I.~Park and E.~Senaha,
  JHEP {\bf 1211}, 116 (2012);
  A.~Merle, V.~Niro and D.~Schmidt,
  arXiv:1306.3996 [hep-ph];
  JHEP {\bf 1309}, 022 (2013)
  [arXiv:1305.3452 [hep-ph]];
  S.~Esch, M.~Klasen and C.~E.~Yaguna,
  arXiv:1308.0951 [hep-ph];
  Y.~Bai and J.~Berger,
  arXiv:1308.0612 [hep-ph].
%
\bibitem{Belanger:2013xza}
  G.~Belanger, B.~Dumont, U.~Ellwanger, J.~F.~Gunion and S.~Kraml,
  arXiv:1306.2941 [hep-ph].
%
%
\bibitem{Dittmaier:2012vm} 
  S.~Dittmaier, S.~Dittmaier, C.~Mariotti, G.~Passarino, R.~Tanaka, S.~Alekhin, J.~Alwall and E.~A.~Bagnaschi {\it et al.},
  arXiv:1201.3084 [hep-ph].
\bibitem{Ade:2013ktc}
  P.~A.~R.~Ade {\it et al.}  [Planck Collaboration],
  arXiv:1303.5062 [astro-ph.CO].
  
  
\bibitem{Knodlseder:2003sv} 
  J.~Knodlseder, V.~Lonjou, P.~Jean, M.~Allain, P.~Mandrou, J.~-P.~Roques, G.~K.~Skinner and G.~Vedrenne {\it et al.},
  Astron.\ Astrophys.\  {\bf 411}, L457 (2003).
 
\bibitem{Su:2010qj} 
  M.~Su, T.~R.~Slatyer and D.~P.~Finkbeiner,
  Astrophys.\ J.\  {\bf 724}, 1044 (2010).
  
\bibitem{Bringmann:2012vr} 
  T.~Bringmann, X.~Huang, A.~Ibarra, S.~Vogl and C.~Weniger,
  JCAP {\bf 1207}, 054 (2012).
  
\bibitem{Adriani:2008zr} 
  O.~Adriani {\it et al.}  [PAMELA Collaboration],
  Nature {\bf 458}, 607 (2009);
  
  \bibitem{AMS}
  AMS Collaboration, M.~Aguilar et al., 
  Phys.\ Rev.\ Lett. {\bf 110}, (2013) 141102.
  
  \bibitem{Hooper:2008kg}
  D.~Hooper, P.~Blasi and P.~D.~Serpico,
  JCAP {\bf 0901} (2009) 025.
  
  \bibitem{Bergstrom:2013jra}
  L.~Bergstrom, T.~Bringmann, I.~Cholis, D.~Hooper and C.~Weniger,
  Phys.\ Rev.\ Lett.\  {\bf 111} (2013) 171101.
  
  \bibitem{Aartsen:2012kia}
  M.~G.~Aartsen {\it et al.}  [IceCube Collaboration],
  Phys.\ Rev.\ Lett.\  {\bf 110} (2013) 131302
  
  \bibitem{indirectDM}
C. Boehm, D. Hooper, J. Silk, M. Casse, and J. Paul, 
Phys.\ Rev.\ Lett. {\bf 92} 101301(2004);
D. Hooper, F. Ferrer, C. Boehm, J. Silk, J. Paul, et al., 
Phys.\ Rev.\ Lett. {\bf 93} 161302 (2004);  
 D. Hooper and T. R. Slatyer, 
arXiv:1302.6589 [astro-ph.HE];
D. Hooper and T. Linden, 
Phys.\ Rev. D {\bf 84} 123005 (2011);  
K. N. Abazajian and M. Kaplinghat, 
Phys.\ Rev. D {\bf 86} 083511(2012).

\bibitem{Buckley:2013bha} 
  J.~Buckley, D.~F.~Cowen, S.~Profumo, A.~Archer, M.~Cahill-Rowley, R.~Cotta, S.~Digel and A.~Drlica-Wagner {\it et al.},
  arXiv:1310.7040 [astro-ph.HE].
 
\bibitem{Angle:2011th} 
  J.~Angle {\it et al.}  [XENON10 Collaboration],
  Phys.\ Rev.\ Lett.\  {\bf 107}, 051301 (2011).
 
\bibitem{Aprile:2012nq} 
  E.~Aprile {\it et al.}  [XENON100 Collaboration],
  Phys.\ Rev.\ Lett.\  {\bf 109}, 181301 (2012);
  E.~Aprile {\it et al.}  [XENON100 Collaboration],
  Phys.\ Rev.\ D {\bf 84}, 061101 (2011);
  E.~Aprile {\it et al.}  [XENON100 Collaboration],
  Phys.\ Rev.\ D {\bf 84}, 052003 (2011).

 
\bibitem{Archambault:2012pm} 
  S.~Archambault {\it et al.}  [PICASSO Collaboration],
  Phys.\ Lett.\ B {\bf 711}, 153 (2012).
  
\bibitem{Behnke:2012ys} 
  E.~Behnke {\it et al.}  [COUPP Collaboration],
  Phys.\ Rev.\ D {\bf 86}, 052001 (2012).
  
\bibitem{Ahmed:2009zw} 
  Z.~Ahmed {\it et al.}  [CDMS-II Collaboration],
  Science {\bf 327}, 1619 (2010);
  Z.~Ahmed {\it et al.}  [CDMS-II Collaboration],
  Phys.\ Rev.\ Lett.\  {\bf 106}, 131302 (2011);
  D.~S.~Akerib {\it et al.}  [CDMS Collaboration],
  Phys.\ Rev.\ D {\bf 82} (2010) 122004.
\bibitem{Agnese:2013lua} 
  R.~Agnese, A.~J.~Anderson, M.~Asai, D.~Balakishiyeva, R.~B.~Thakur, D.~A.~Bauer, J.~Billard and A.~Borgland {\it et al.},
  arXiv:1309.3259 [physics.ins-det].
  
\bibitem{Agnese:2013rvf} 
  R.~Agnese {\it et al.}  [CDMS Collaboration],
  [arXiv:1304.4279 [hep-ex]].
  
    
 
  
\bibitem{Aalseth:2012if} 
  C.~E.~Aalseth {\it et al.}  [CoGeNT Collaboration],
  Phys.\ Rev.\ D {\bf 88}, 012002 (2013); 
  C.~E.~Aalseth, P.~S.~Barbeau, J.~Colaresi, J.~I.~Collar, J.~Diaz Leon, J.~E.~Fast, N.~Fields and T.~W.~Hossbach {\it et al.},
  Phys.\ Rev.\ Lett.\  {\bf 107}, 141301 (2011).
  
\bibitem{Bernabei:2013cfa} 
  R.~Bernabei, P.~Belli, S.~d'Angelo, A.~Di Marco, F.~Montecchia, F.~Cappella, A.~d'Angelo and A.~Incicchitti {\it et al.},
  Int.\ J.\ Mod.\ Phys.\ A {\bf 28}, 1330022 (2013);
  R.~Bernabei,
  Annalen Phys.\  {\bf 524}, 497 (2012).
 
\bibitem{Gresham:2013mua} 
  M.~I.~Gresham and K.~M.~Zurek,
  arXiv:1311.2082 [hep-ph].
  
\bibitem{Akerib:2013tjd}
  D.~S.~Akerib {\it et al.}  [LUX Collaboration],
  arXiv:1310.8214 [astro-ph.CO].
  
\bibitem{DelNobile:2013gba} 
  E.~Del Nobile, G.~B.~Gelmini, P.~Gondolo and J.~-H.~Huh,
  arXiv:1311.4247 [hep-ph].
  
\bibitem{Clowe:2006eq}
  D.~Clowe, M.~Bradac, A.~H.~Gonzalez, M.~Markevitch, S.~W.~Randall, C.~Jones and D.~Zaritsky,
  Astrophys.\ J.\  {\bf 648} (2006) L109.
  
\bibitem{Bento:2000ah} 
  M.~C.~Bento, O.~Bertolami, R.~Rosenfeld and L.~Teodoro,
  Phys.\ Rev.\ D {\bf 62}, 041302 (2000);
  J.~McDonald, N.~Sahu and U.~Sarkar,
  JCAP {\bf 0804}, 037 (2008).
 
\bibitem{Arrenberg:2013rzp}
  S.~Arrenberg, H.~Baer, V.~Barger, L.~Baudis, D.~Bauer, J.~Buckley, M.~Cahill-Rowley and R.~Cotta {\it et al.},
  arXiv:1310.8621 [hep-ph].

\bibitem{projectX} A.~S.~Kronfeld {\it et al.},
  arXiv:1306.5009 [hep-ex].
 
  
\end{thebibliography}
\end{document}